# Properties of Liquid Crystalline Elastomer Foams

*Oliver Dai* [a], *Andrew Terentjev* [b], *and Eugene M. Terentjev*\*[a]

(a) Cavendish Laboratory, University of Cambridge, Cambridge CB30US, U.K.
   E-mail: emt1000@cam.ac.uk
(b) Cambridge Smart Plastics, Cambridge CB43RH, U.K.

**Abstract.**
We investigate how controlled foaming alters the mechanical dissipation of liquid crystalline elastomers (LCEs). Using thermally expandable microspheres, we generate homogeneous foams with precisely tuned bubble volume fractions up to ~13%, and compare their behavior with non-mesogenic silicone analogues. We show that microsphere expansion induces a particle-centered mesogenic interphase, arising from local elastic distortion and preferential alignment of mesogenic units at the inclusion surface. At low bubble volume fractions (≈0.5–5%), these interphases remain spatially isolated and produce a pronounced, non-monotonic enhancement of damping, with the loss factor reaching *tan*δ ≈ 0.2 even in the isotropic regime. At higher loadings, interphase overlap and mechanical constraint suppress this effect, and the dissipation returns towards baseline elastomeric values. Large-strain tensile tests and impact experiments exhibit the same non-monotonic trend, demonstrating that low-density LCE foams achieve the highest mechanical energy absorption per unit mass. Compared with conventional high-porosity polymer foams used for acoustic damping, these materials retain sufficient mechanical integrity to sustain impact loads, establishing a microstructural route to engineer high-performance damping in soft solids.

## 1. Introduction

Liquid crystalline elastomers (LCEs) combine long-range orientational order with rubber elasticity, giving rise to a unique coupling between molecular alignment and macroscopic mechanics.[1,2] This interplay underpins several hallmark properties, including large thermally driven actuation strains,[3] soft elasticity,[4,5] and anomalously high mechanical dissipation in the nematic phase.[6] While actuation and soft elasticity have dominated both fundamental studies and applications of LCEs, anomalous damping has more recently emerged as a property of exceptional technological promise, with relevance to impact mitigation, vibration control, adaptive interfaces, and pressure-sensitive adhesion. Until recently, there was no clear understanding that the anomalous damping in nematic LCE is an effect completely separate from the first two. There is a view that anomalous LCE damping (and the closely associated adhesion) has a greater application potential than any other property of these materials.[2]
The enhanced dissipation of nematic LCEs originates from director fluctuations that are elastically coupled to the polymer network but relax through distinct molecular pathways, providing an internal channel for mechanical energy loss that is absent in conventional elastomers. Considerable effort has been devoted to amplifying this effect by molecular design strategies that increase orientational mobility or introduce dynamic mesogenic segments. In parallel, however, structural approaches to tuning dissipation — widely exploited in conventional polymers through foaming — remain largely unexplored in LCEs.

Foaming is a classical route to modify polymer mechanics, enabling reductions in density, tailored stiffness, and enhanced energy absorption through controlled microstructural heterogeneity.[7] In conventional elastomers, the introduction of voids or inclusions generates complex stress fields, promotes localized deformation, and increases internal friction, often leading to improved damping.[8,9] Despite its technological maturity in polymer engineering, this strategy has only recently been applied to LCEs,[10,11] primarily in the context of soft actuation, biocompatible scaffolds and gripping, leaving the fundamental implications for damping and dissipation largely unaddressed. In contrast to these approaches, our work exploits low-volume-fraction foaming to deliberately induce a mesogen-aligned interphase and thereby amplify intrinsic dissipation in LCEs, even in the isotropic regime.

Here we employ a new method, utilizing the 'Expancel microspheres' (ECM) technology of Nouryon (http://www.expancel.com): thermoplastic microspheres enclosing hydrocarbon fluid expand when heated and the fluid evaporates, creating thin bubbles of controlled diameter. We prepare the homogeneous foams of a standard mainstream thiol-acrylate nematic LCE using the ECM bubbles of uniform 20 μm size and increasing fraction of gas phase. In this way, we demonstrate that controlled foaming of LCEs using thermally expandable microspheres provides a powerful route to engineer mechanical damping. We uncover a non-monotonic dependence of dissipation on inclusion concentration, with a pronounced maximum at low loadings. Combining optical microscopy, tensile testing, dynamic mechanical analysis, and impact experiments, we show that microsphere expansion induces a particle-centered mesogenic interphase that governs mechanical response in both nematic and isotropic regimes. This establishes a unified microstructural framework linking local elastic distortion, mesogen alignment, and macroscopic energy dissipation in foamed liquid crystalline elastomers.

## 2. The Range of LCE Foams

LCE foam samples were prepared by incorporating a uniform distribution of unexpanded ECM into the reaction mixture and heating to expand the ECM before LCE curing. The discussion and the details of the ECM type we chose to use are in the "Methods" section below. Briefly, we use the 920DU40 grade ECM, with an approximate outer radius of 6 μm in the initial compressed state and of 20 μm in the expanded state. There is a broad selection of ECM grades from their manufacturer, and our choice was dictated by the following factors: the thinnest plastic 'skin' after the gas expansion (to introduce the minimal mechanical effect from it), the lowest temperature of expansion (to reduce the catalyst evaporation during the expansion phase), and the biggest size of expanded 'bubbles': optimizing between these factors has dictated the choice of 920DU40 grade. The knowledge of these dimensions allows us to recalculate the volume fraction of gas bubbles in the elastomer matrix from the initial value of wt% ECM added into the reaction resin, given in **Table 1**.

We also considered a reference material: the standard silicone (PDMS) prepared with several ECM concentrations to model the low-concentration foams, with density similar to their LCE analogues, and also the commercial product 'Alpha Gel' ($\alpha$-Gel) produced by Taica Corporation (https://taica.co.jp/gel/en/product/), which is a silicone foam with about 70% concentration of bubbles (i.e. low overall density). We did not have an LCE with similarly high concentration of air bubbles, but we reasoned that at such a low fraction of polymer the anomalous dissipation properties of LCE would not contribute much – the damping in such low-density foams is mostly due to structural composition of the metamaterial. Therefore, we considered the properties of Alpha Gel a model for the high-concentration LCE foam. Table 1 gives the information about each foam composition, and also their mass density. The knowledge of that is important when we are comparing different foams: later on we will see how different-density foams perform in impact damping in a fixed-thickness pad. However, we may also want to normalize different foams to the same mass of polymer that is meeting the impact (by increasing the pad thickness): in such an adjusted (scaled) form, the results will specifically address the effect of internal metamaterial structure without being dominated by the overall weakness of the foam.

ECM gas expansion is limited by the stiffness of the matrix in which it expands; when we attempted heat treatment of ECM post-crosslinking, we observed insufficient ECM expansion: the elastomer matrix was offering sufficient resistance to the attempted gas expansion. Expansion of ECM at the very early stages of elastomer crosslinking, as described in the "Methods" section 6, proved more effective and produced gas bubbles of expected radii, as shown below in **Figure 1**. ECM volume fractions of up to 13% were investigated in this study; higher volume fractions were not of interest, as it was expected that at high concentrations the unique liquid-crystalline effects would diminish and the mechanical properties would approach those of an ideal elastomeric foam composite. The production protocol was optimized to maintain a homogeneous distribution of ECM within the sample. In spite of our thorough mixing of ECM in the resin, the particle distribution was never truly homogeneous (see Figure 1). There were inevitable fluctuations in the bubble density, but the largest issue was the floating up of low-density expanded ECM to the surface in the low-viscosity matrix that had not yet been cured. To address this, we increased the catalyst concentration so that sufficient gelation occurs in the short time of high-temperature exposure for the ECM expansion and minimizes the bubble floatation. The same principles were applied to the PDMS mixtures with ECM and crosslinking them into elastomeric foams.

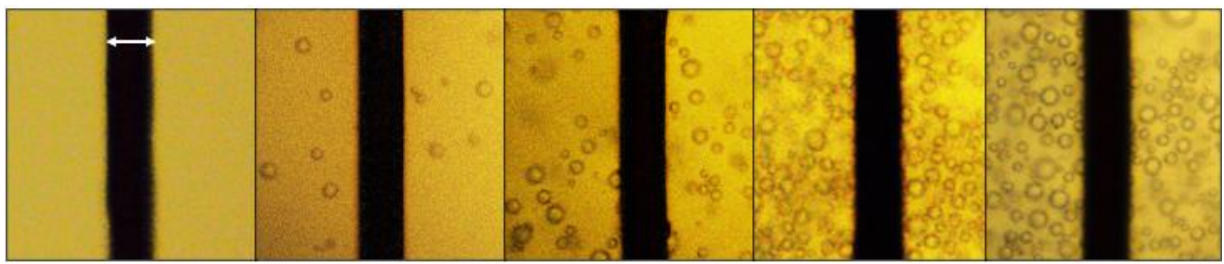

**Figure 1.** Photos of LCE with expanded ECM. The dark bar in every image is the human hair, of 80 μm diameter, providing the reference length scale. The sequence of images has 0%, 0.25%, 2.5%, 4.7% and 13% volume fraction of gas bubbles in the matrix (see Table 1 and the details of preparation in the "Methods" section 6).

**Table 1.** The list of LCE foams prepared for the final presentation, and the reference PDMS foams. Their label (such as 'LCE x') reflects the volume fraction x% of expanded bubbles in the matrix, while the second column shows the weight fraction of ECM to be added to the reacting resin mixture. The right columns show the <u>mass density</u> of each foam material, and the <u>equivalent thickness</u> of the pad to have the same mass of polymer in the impact path.

| Sample label | ECM weight fraction [%] | Expanded ECM volume fraction [%] | Density (g/cm$^3$) | Equivalent thickness (mm) |
|---|---|---|---|---|
| LCE 0 | 0 | 0 | 1.30 | 3 |
| LCE 0.25 | 0.004 | 0.25 | 1.297 | 3.01 |
| LCE 0.6 | 0.01 | 0.62 | 1.292 | 3.02 |
| LCE 1.2 | 0.02 | 1.2 | 1.284 | 3.04 |
| LCE 1.8 | 0.03 | 1.8 | 1.277 | 3.06 |
| LCE 2.5 | 0.04 | 2.5 | 1.268 | 3.08 |
| LCE 3.6 | 0.07 | 3.6 | 1.253 | 3.11 |
| LCE 5 | 0.09 | 4.7 | 1.239 | 3.15 |
| LCE 9 | 0.17 | 9.0 | 1.183 | 3.30 |
| LCE 13 | 0.26 | 13 | 1.131 | 3.45 |
| PDMA 0 | 0 | 0 | 0.97 | 3 |
| PDMS 1.5 | 0.025 | 1.5 | 0.955 | 3.05 |
| PDMS 5 | 0.1 | 5 | 0.922 | 3.16 |
| PDMS 10 | 0.22 | 10 | 0.873 | 3.33 |
| Alpha Gel | n/a | 75 | 0.242 | 12 |

## 3. Nematic Order and Equilibrium Mechanical Properties

Spherical ECM inclusions affect the mechanical properties of LCE on multiple levels. Modelling the ECM as simple air bubbles, the Eshelby theory [12] can be used to predict the evolution of the Young Modulus of the composite material. However, in addition, surface molecular-scale interactions between the acrylate shell and liquid-crystalline polymer chains lead to both filler particle effects and local changes in the LCE nematic order. **Figure 2** shows the macroscopic optical effect of this changing nematic order. With increasing ECM volume fractions from 0% to 1%, the LCE samples grow more transparent, indicating reduced scattering and haze of the natural polydomain nematic LCE. This effect is known in all partially ordered systems: Bernland et al.[13] have shown that the creation of small-scale structures reduces haze in partially crystalline structures like polypropylene. A similar effect arises here, where ECM addition to LCE leads to regions of frustrated orientational order around ECM-LCE interfaces and thus reduced haze. With increasing ECM populations, the disordered zone volume fraction increases, the overall average nematic order decreases and further scattering suppression occurs. The macroscopic LCE structure remains polydomain, since the characteristic domain size in LCE is just 1-2 µm [14,15]: on stretching to >100% strain, the samples grow transparent on polydomain-monodomain transition,[16] and recover the scattering polydomain texture on annealing to isotropic phase.

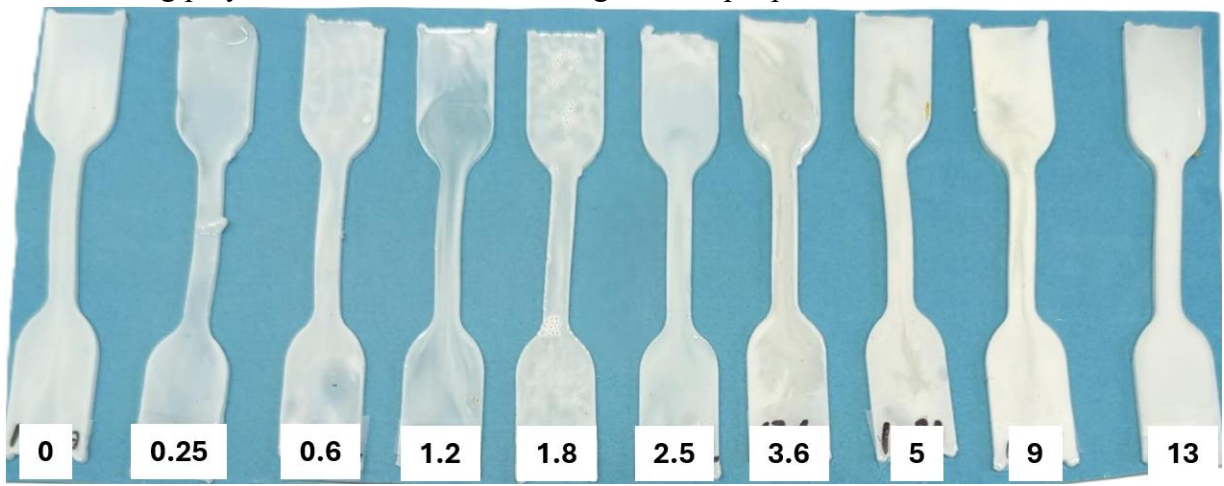

**Figure 2.** Photos of ASTM D638 samples of LCE foams with ECM bubbles volume concentration increasing from zero up to 13%. The natural polydomain nematic LCE state is highly scattering light (hence white color). As the bubbles (impurities) increase, the order decreases and the LCE becomes more transparent. As the bubbles saturate the rubber, the light scattering increases again, now dominated by the metamaterial structure.

At larger ECM populations, beyond 2-3 vol%, the scattering effect of the air bubbles dominates, and sample opacity increases once more. This is supported by the retained opacity of these samples when stretched to >100% strain, or annealing to isotropic phase, showing that enhanced scattering effects are due to the gas bubbles in the matrix.

The homogeneity of ECM distribution through the sample thickness was a major challenge for us, because in a low-viscosity pre-crosslinked resin the expanded bubbles were aggregating at the top due to buoyancy. We have developed a protocol of optimization, described in the "Methods" section 6, aiming to prevent this segregation. However, full homogeneity of low-concentration LCE foams could never be achieved: fluctuations of the bubble density were relevant on the mm-scale of testing samples, as shown in Figure 1 above. The mechanical properties of individual dogbone samples can be dominated by regions of local weakness caused by local increase in ECM density across the thin (stretching) section of the dogbone. Local inhomogeneity can also lead to overlapping regions of locally aligned mesogenic units between close-by ECM, disrupting mechanical property consistency. This effect is most prevalent at low ECM fraction: pure LCE samples produce very consistent stress-strain curves, as do the LCE foams with higher-concentration of ECM bubbles.

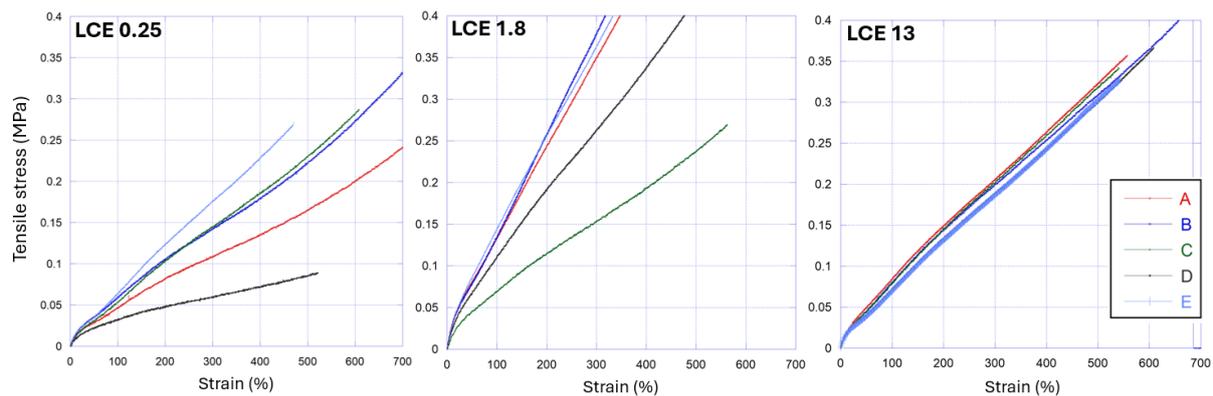

**Figure 3.** Tensile tests illustrating the variability of samples, for low/ medium/ high ECM loading (as labelled in the plots). Each plot has 5 different samples (curves A to E in each plot) of exactly the same preparation. The high variability at low and medium ECM concentration is due to randomly inhomogeneous distribution of bubbles across the testing zone of the dogbone. At high ECM loading, LCE 13, the role of inhomogeneity is reduced.

**Figure 3** illustrates this effect of density fluctuations affecting the tensile mechanical response in the samples with the ASTM D638 standard cross-section dimensions of 6x3mm. Given the variability of stress-strain curves between low ECM-loading samples, average curves were taken from many repetitions. The results, however, that increased ECM-loading lead to a visible rise, and then fall in stored energy under the corresponding stress-strain curve – the superposition of which is presented in **Figure 4**.

Notably, the characteristic soft elasticity plateau of pure LCE is suppressed even at marginal ECM addition (0.6 vol%). Soft elasticity in pure LCE arises due to nematic director reorientation in the domains, which occurs with minimal resistance,[1,4,5] but when mesogenic units are preferentially aligned at ECM surfaces, elastic deformation requires substantially more energy. We have demonstrated that even slight ECM addition therefore suppresses the soft elasticity plateau.

At low ECM concentration, there is a stark rise in stored energy with ECM population, peaking at around 3-4% volume fraction. Such stiffening originates from LCE-ECM surface interactions that drive preferential LCE mesogen alignment near the surface, producing a paranematic phase there. This effect is similar to a nano-filler in rubbers; however, the trend

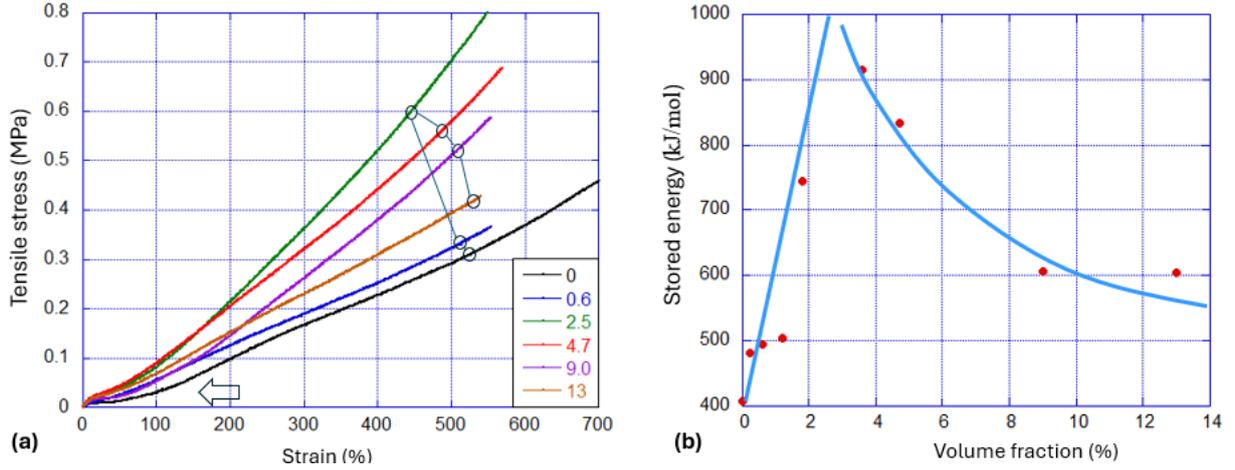

**Figure 4.** Tensile tests: (a) Overlaid stress-strain curves of several LCE foams with the volume fraction of expanded bubbles changing from zero (in the base LCE control) up to 13%. Notably, all curves are above the reference LCE response, and the change is non-monotonic: the stiffest elastomer is at 2.5% foam volume. (b) The plot of Stored Work (toughness) points calculated from the tensile curves up to 400% strain. This plot makes the non-monotonic change in foamed LCE properties much more visible. The lines in this plot represent the two regimes: of low- and high- concentration of bubbles, Eqs. (1-2).

rises far faster than that predicted by the Einstein-Guth-Gold relation for the Young modulus:[17]

$$E_c = E(1 + 2.5\phi + 14.1\phi^2). \qquad (1)$$

This is expected due to the highly non-linear nature of the LCE stress-strain response, where the Young modulus and stored energy cannot be simply correlated. Moreover, the Einstein relation, based on the calculation of local internal stress around spherical inclusions, does not take into account the deformable nature of the ECM, observed to deform beyond 300% lengthwise under extensive tensile strain, considerably enlarging the paranematic phase zone, see **Figure 6** below, and Supporting Information S3.

At higher ECM concentration, the stored energy drops off again, following the Eshelby inclusion theory.[12] Although Eshelby's theory ignores the inclusion-matrix surface energy, Style et al.[18] have extended it to show that in the limit of small surface tension $\gamma$, $R \gg \gamma/E$, where $R$ is the inclusion radius and $E$ the Young modulus of the matrix, Eshelby's result for composite stiffness $E_C$ is recovered:

$$E_C = E \frac{1}{(1+5\phi/3)} . \qquad (2)$$

The surface energy of the Expancel acrylate plastic is known to be $\gamma \sim 50$ mN/m (see Figure S2 Supporting Information), and the Young modulus of LCE has been measured as around $E \sim 50$ kPa, yielding $R \cdot E \sim 1$ N/m $\gg \gamma$, and hence the Eshelby equation is fully applicable. The decrease in stored energy seen in Figure 4(b), which is proportional to the elastic modulus, agrees with this prediction. In this region of higher ECM bubbles concentration, the effect on LCE microstructure is marginal and the properties approach those of an ideal foam composite. The higher-concentration part of the plot in Figure 4(b), reflects the inverse

dependence on bubble fraction in Equation (1), and also matches the results reported by Style et al.[18]

## 4. Dynamic Mechanical Properties — Mechanistic Interpretation

DMA characterization is important to understand the microscopic properties of our inhomogeneous metamaterials, both LCE foams and PDMS foams at the same concentration, examined for reference. LCE materials undergo two important phase transitions visible in a DMA scan: a lower-temperature glass transition, appearing as the tall peak centered at Tg, and the nematic-isotropic transition at a higher temperature $T_{NI}$. Often, these two transitions are also identified in the Differential Calorimetry (DSC) scans, however, our highly inhomogeneous 'metamaterials' do not show good DSC peaks. Therefore, the DMA scans are our only way to quantitatively identify the transition points. The isotropic-nematic transition signature in DMA appears both in the storage modulus and in the loss factor *tan*δ. The storage modulus has a characteristic minimum at this transition,[1,6] and then starts slow rising showing the isotropic rubber elasticity. The loss factor is characteristically high in the nematic phase (the phenomenon of 'anomalous dissipation') and then drops to a low constant value in the isotropic phase. Therefore, a combination of a glass transition peak, and an enhanced dissipation plateau produces the characteristic shape of *tan*δ vs. T with a 'shoulder' in the nematic phase. **Figure 5** gives the overlaid plots of the loss factor *tan*δ against temperature, taken at a constant frequency of 1Hz. We plot only the loss factor, since on increasing foam concentration both storage and loss moduli will diminish strongly (simply because there is less polymeric material tested), but their ratio *tan*δ remains a representative intensive characteristic. Three different regions in that plot are zoomed-in in the panels (b,c,d). In all three cases, there is a clear trend that that low-density foams (0.25, 0.6 and 2.5%) have a higher loss factor than either pure LCE or high-density foams.

The important and striking difference occurs in the isotropic regime of the LCE foams, **Figure 5(b)**, where long-range orientational order is absent and the matrix would ordinarily behave as a simple isotropic rubber. While both the base LCE and the PDMS control foams exhibit low, nearly composition-independent loss factors in this regime, low-volume-fraction LCE foams display a pronounced enhancement of dissipation, with tan δ reaching ~0.2–0.25 at bubble fractions of 0.5–5%. This non-monotonic response closely mirrors the trend observed in tensile energy storage (Figure 4), indicating a common microstructural origin rather than a trivial density, void, or shell effect.

Such behavior cannot be explained by classical foam mechanics, gas-phase dissipation, or deformation of the microsphere shell, all of which would be expected to produce a monotonic dependence of damping on porosity and are observed in the PDMS control foams. The absence of a comparable enhancement in these chemically similar but non-mesogenic systems directly implicates liquid-crystalline order, rather than generic stress or shell effects, as the dominant source of dissipation. Instead, it points to the emergence of a particle-centered interphase in which elastic distortion of the network locally aligns mesogenic units, generating a zone of constrained polymer dynamics and enhanced internal friction. In the nematic state, this interphase manifests as a birefringent halo surrounding each microsphere, while above $T_{NI}$ it persists as a dynamically ordered paranematic region that continues to dissipate mechanical energy.

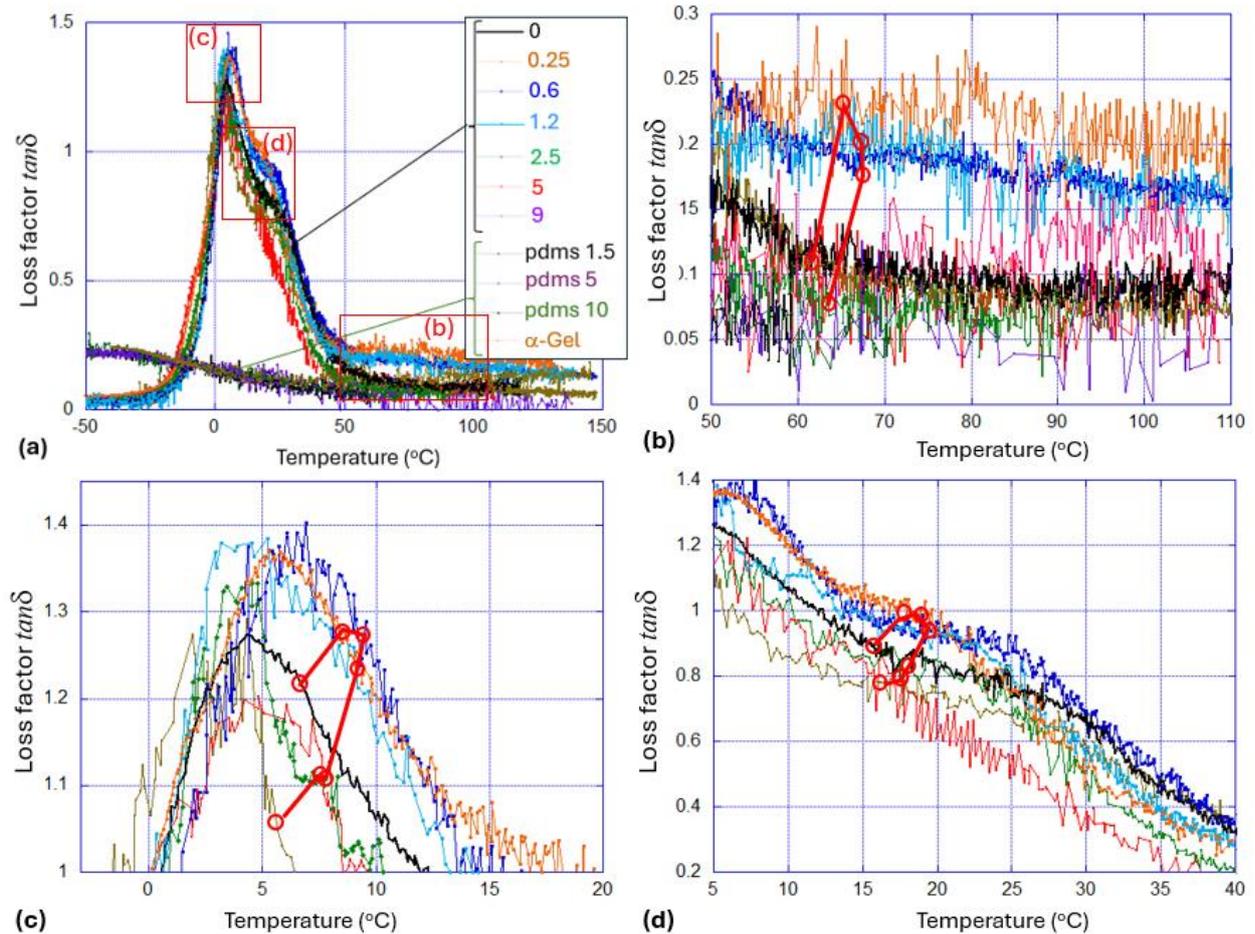

**Figure 5.** (a) Overlaid DMA results: $tan\delta$ curves for all our LCE foams, and also for the reference PDMS (silicone foams). In each case the number in the legend refers to the volume fraction of bubbles in the elastomer, same as in Table 1. Also plotted the $tan\delta$ of α-Gel, which is remarkable similar to the denser PDMS materials, because the loss factor ratio $tan\delta$ reflects the normalized properties of the polymer. We zoom into three characteristic regions: (b) The section of $tan\delta$ curves in the isotropic phase of LCE, where we see that low-density foams (0.25, 0.6 and 2.5%) have a much higher loss factor compared to all other materials, (c) where we focus on the glass transition peak, and (d) on the region of enhanced damping in the nematic phase. In all cases, the non-monotonic behavior, with low-density foams having increased loss, agrees with the trend seen in Figure 4(a).

Direct optical evidence for this interphase is provided by polarized microscopy (**Figure 6**). Importantly, the birefringent signal is spatially localized around individual inclusions and rotates systematically with analyzer orientation, distinguishing elastically coupled mesogen alignment from passive stress birefringence of an isotropic matrix. Even in the isotropic phase, a pronounced birefringent corona is observed around individual expanded microspheres, demonstrating that elastic stretching of the network stabilizes local mesogenic alignment above the bulk transition temperature. In the nematic phase, this aligned zone extends to approximately one to two particle radii from the inclusion surface, defining a finite interphase volume. At low microsphere concentrations, these interphases remain spatially isolated and act as independent dissipative units, leading to a maximum in macroscopic damping. As the inclusion density increases beyond ~5 vol%, neighboring interphases begin to overlap and mechanically constrain one another, reducing local strain amplitudes and suppressing their dissipative efficiency.

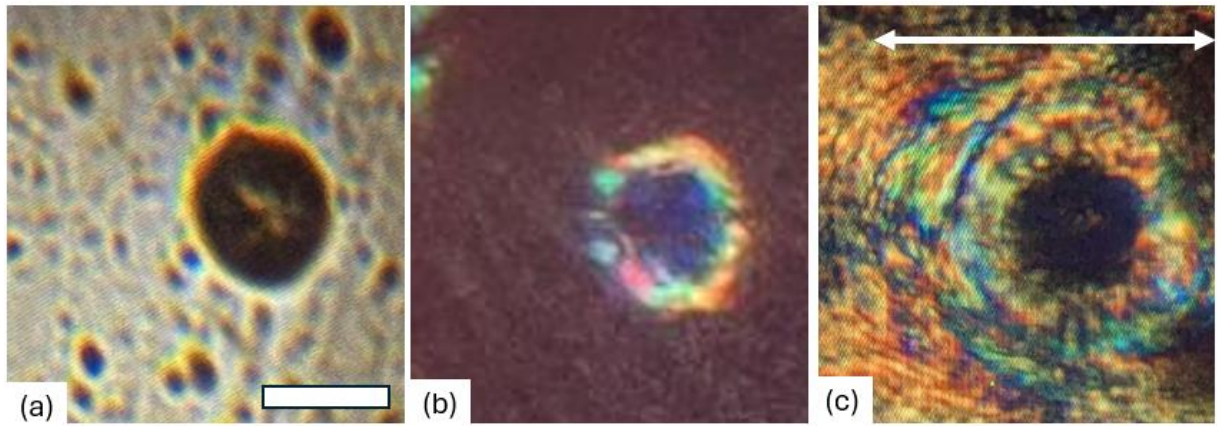

**Figure 6.** Optical microscope images of an example expanded ECM: (a) in bright field, with the scale bar =20μm, (b) between crossed polarizers in the isotropic phase, highlighting the paranematic order around the stretched bubble surface, and (c) between crossed polarizers in the polydomain nematic phase, having the whole matrix birefringent, but still leaving the locally stretched and aligned region clearly visible (marked by the arrow).

Together, the optical and dynamic mechanical results establish a unified microstructural framework for dissipation in foamed LCEs. Microsphere expansion generates localized elastic distortion that induces mesogenic alignment and creates a finite dissipative interphase. This interphase governs mechanical damping in both the nematic and isotropic regimes, producing a pronounced non-monotonic dependence of tan δ on bubble volume fraction and linking local director organization directly to macroscopic energy dissipation.

PDMS has a very low Tg, and no anomalous effects due to the orientational order of LCE, so the loss factor of its foams is consistently low, and shows no difference between samples at such a low concentration. Remarkably, the low-concentration foams, and the 75% void a-Gel have exactly the same *tanδ*, reflecting the silicone material (as do the LCE foam curves, where the small change is only due to the shifting nematic-isotropic transition $T_{NI}$).

To explore the role of paranematic phase above $T_{ni}$ we monitored the birefringence in the materials. **Figure 6** shows the same example ECM bubble observed in the bright field (image a), and between crossed polars to expose birefringence: image (b) in the isotropic phase and image (c) in the nematic phase, below $T_{ni}$. It is clear that the locally stretched 'skin' of elastomer around the bubble retains the orientational order above $T_{ni}$. The image (c), below $T_{ni}$, helps visualize the extent of this local orientation caused by elastomer stretching: this region (where the Eshelby deformations occurs [12]) spans about twice the size of the bubble itself.

## 5. Impact Damping — Design-Driven Interpretation

To evaluate the practical implications of microsphere-induced dissipation, we performed controlled drop-weight impact tests on foamed LCE pads and reference silicone foams. When pads of identical thickness (of 3mm) are compared under the same impact of 5J, **Figure 7(a),** weak high-porosity foams fail catastrophically under impact, transmitting forces comparable to direct metal–metal contact and providing negligible protection. In contrast, low-density LCE foams with moderate bubble fractions withstand impact without structural collapse, limiting peak transmitted forces to ~2–3 kN and demonstrating robust energy absorption.

To disentangle intrinsic dissipation from trivial density effects, impact experiments were also conducted under mass-normalized conditions. With the knowledge of foam densities, listed in Table 1, we are able to calculate how thick the damping pad should be for the impact to encounter the same amount of polymeric material: the lower the density, the greater the pad thickness for this normalized comparison. So, the pad thickness was adjusted such that the same total polymer mass lay in the impact path, **Figure 7(b)**. Under these conditions, low-density LCE foams significantly outperform both unfoamed LCE and PDMS foams, producing lower peak forces and broader impulse profiles at identical polymer mass, confirming that the enhanced impact mitigation arises from intrinsic dissipation rather than geometric cushioning or reduced stiffness. This behavior directly reflects the enhanced internal dissipation arising from the microsphere-induced interphase, rather than simple geometric cushioning. Notably, foams with bubble fractions near 5–10% exhibit the optimal balance between mechanical integrity and dissipative efficiency, identifying a practical design window for impact-mitigating LCE metamaterials.

These results highlight a critical distinction between structural foams optimized for acoustic damping and mechanically robust metamaterials capable of sustaining impact loads. While extremely high porosity maximizes compressibility, it severely compromises load-bearing capacity. In contrast, moderate microsphere loading preserves elastomer continuity while introducing a highly dissipative microstructure, enabling superior impact protection at substantially reduced density.

Since the width of the impact peak increases with the foam thickness, the peak force gets lower because of the momentum-conservation constraint: $Mv = \int F(t)dt$. Therefore, we must conclude that the low-density LCE foams offer relatively greater protection to the target: the peak force reaches just 2.2 kN for the LCE 9 material. As expected, LCE material offers a much greater impact damping compared to the matching PDMS elastomer, but the case of weak α-Gel is special and worth examining because we expect the high-porosity LCE foam will perform similarly. In this test, we used a stack of 8x 3mm α-Gel pads, 24mm thickness in total, and the force peak was much delayed. In this normalized test where we probe the material, the peak force is like that of PDMS-ECM foams. However, despite this thickness, the damping pad was still broken through, see **Figure 8**.

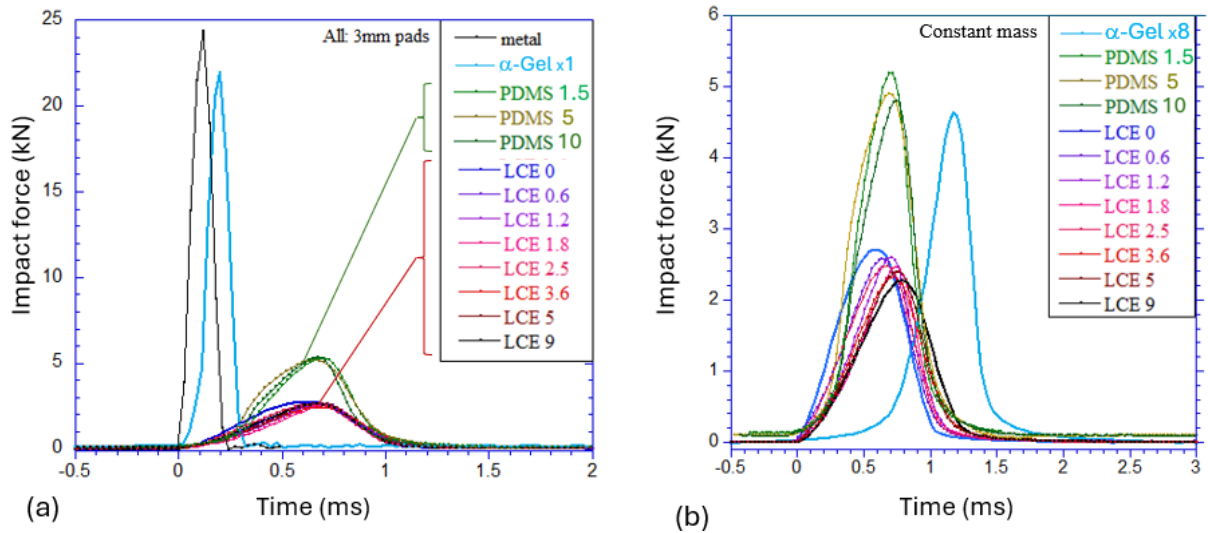

**Figure 7.** (a) All test results of 5J impact on 3mm-thick pads overlaid, with a curve for impact on bare metal also given for reference. The curve 'α-Gel x1' refers to a single 3mm-thick pad of α-Gel. (b) Comparison of impact curves, with the pad thickness adjusted for the decreasing density of the foam (see text for detail). The curve 'α-Gel x8' refers to the 24mm thick α-Gel (8-pad stack), the damaging outcome of this impact illustrated in Figure 8 below.

We may expect that between the last (LCE 9) curve in Figure 6(b) and the α-Gel x8 curve, there would be an optimal foam concentration in LCE where the normalized damping would be the greatest. It is likely that 9-10% of the bubble volume is close to that optimum where the enhanced dissipation due to the added microstructure is still supported by the sufficient strength of the metamaterial; clearly at 75% void in α-Gel this strength is too low for any protection.

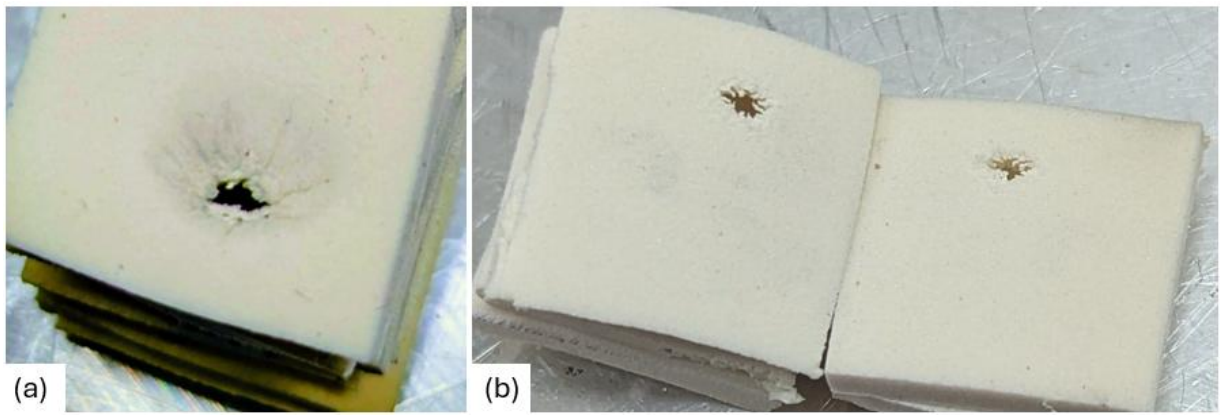

**Figure 8.** (a) The full 8-stack of 3mm pads after the impact has significant damage. (b) The break-through is evident in the middle of the stack.

## 6. Conclusion — Design Principles for Dissipative Soft Solids

The anomalous damping behavior of LCE materials is increasingly drawing attention for its diverse application potential, and further refinement of this behavior represents an important academic direction. This may be achieved by playing with base polymer properties, but in this study, we focus on the mechanical alteration of LCE damping behavior via foaming, minimizing the impact on polymer composition. We have demonstrated that controlled foaming of liquid crystalline elastomers using thermally expandable microspheres provides a powerful microstructural route to engineer mechanical damping. Microsphere expansion generates localized elastic distortion that induces mesogenic alignment and forms a finite dissipative interphase, which governs mechanical response in both nematic and isotropic regimes. This produces a pronounced, non-monotonic enhancement of dissipation at low bubble volume fractions, yielding *tanδ* values approaching 0.2 even in the isotropic state.

By linking microstructural organization directly to macroscopic energy loss, this work establishes a clear design principle: maximum damping in soft solids is achieved not through extreme porosity, but through the controlled introduction of localized, elastically coupled interphases that concentrate internal friction while preserving mechanical integrity. In LCE foams, this balance is optimized at moderate microsphere loadings, enabling exceptional impact mitigation at substantially reduced material density.

Beyond liquid crystalline elastomers, the interphase-engineering strategy demonstrated here offers a general framework for designing high-performance damping materials. By exploiting elastic distortion fields around inclusions to generate dynamically ordered zones, it should be possible to extend this concept to a wide range of responsive polymers, soft composites, and mechanical metamaterials, opening new routes toward lightweight, adaptive, and highly dissipative soft matter systems.

A few factors warrant future study. Studies on the microstructure evolution of paranematic interphase, as well as separating the alignment effect from pure foaming, are necessary. The in-application damping performance of these materials, including their cyclability, also needs further characterisation. This includes overcoming the polymer synthesis problem of distribution inhomogeneity, which produces local weakness and hence sample-to-sample tensile property inconsistencies.

## 7. Experimental Methods

*Synthesis of LCE*
Thiol-acrylate main-chain LCE at 10% crosslinking density was prepared following a modification of standard methods reported previously [19] with a single-step crosslinking reaction of thiol-acrylate Michael addition. The diacrylate reacting monomer, 1,4-bis-[4-(3-acryloyloxypropypropyloxy) benzoyloxy]-2 methylbenzene (RM257), was purchased from ChemFish Tokyo Co. Ltd, the dithiol spacer, 2,2'-(ethylenedioxy) diethanethiol (EDDT) and tetrathiol crosslinker pentaerythritol tetrakis (3-mercaptopropionate) (PETMP), were purchased from Sigma Aldrich. Toluene (Fisher Chemical) was used as a solvent to control the reaction rate. Triphenylphosphine (TPP, Sigma Aldrich) was used as the catalyst of the thiol-acrylate Michael addition reaction. As the radical scavenger, butylated hydroxytoluene (BHT, from Sigma Aldrich) was used to suppress the unwanted radical polymerization reaction between acrylates. All chemicals were used in their as-received condition with no purification (see Figure S1: Supporting Information). RM257 (16 g), BHT (0.32 g), and 1 g of Toluene were mixed and then heated at T≈85°C for 30 min. EDDT (4.46 g) and PETMP (0.66 g) were then added, and TPP (0.8 g) was added to start the Michael-addition reaction between thiol and acrylate groups. The mixture was mixed and transferred to a Teflon mould and heated at T≈160°C for 8 min, before moving to room temperature for overnight curing.

*Making a foam with ECM*

The structure of the unexpanded ECM particles (920DU40 grade, received from Nouryon) is known from the manufacturer's data (see Figure S2: Supporting Information). The particles consisted of a liquid blowing agent (average inner radius of its spherical drop 6 μm) coated by a thermoplastic shell of thickness 2 μm. On heating at T≈160°C for 8 min, the liquid rapidly evaporates, and the ECM sphere expands into a thin shell of thickness 0.1 μm and average radius 20 μm: an increase of 4x in radius or 64x in volume. According to the manufacturer Nouryon, the thin shell is an acrylate copolymer, with a density of 1.1 g cm$^{-3}$.

The overall density of a single unexpanded and expanded ECM, $\rho_i$ and $\rho_f$ respectively, was estimated using the data provided as
$$\rho_i = 0.774 \text{ g cm}^{-3}$$
and
$$\rho_f = 0.021 \text{ g cm}^{-3}$$
The weight fraction of ECM, *M*, is related to the observed ECM volume fraction, $R_V$, via:
$$R_V = \frac{1}{1 + \rho_{ECM}/(\rho_{LCE} \times M)}$$
for $\rho_{LCE}$ and $\rho_{ECM}$ the LCE and post-expansion ECM densities, respectively. Proof of these results is given in Supporting Information Section S4.

Desired amounts of ECM were added directly after the addition of PETMP to achieve the required air volume fractions (Table S5, Supporting Information). The mixture was first homogenised under vacuum using a Hauschild DAC M-series SpeedMixer for 1 min, then 0.8g of TPP catalyst was added, and the mixture was mixed again for another 1 min. The mixture was then heated to 160°C for 8 minutes to activate ECM expansion. This stage has dictated the choice of TPP catalyst, with a boiling point of 377°C, so it was not affected by this heating. The other precursors also remain stable at this temperature.

An amount of TPP, a high expansion temperature, and an expansion time of 8 minutes were used to maximise the rate of in-situ and in-mould LCE curing whilst allowing the ECM to expand to the desired radius. It was found that under standard protocols with slower curing, where the mixture retains low viscosity for longer times, very low-density expanded ECM 'bubbles' strongly segregate to the top of the curing samples by buoyancy. Our protocol optimises the through-thickness ECM homogeneity and their uniform size distribution. The final ECM volume fractions are given in Table 1.

*Synthesis of PDMS*
PDMS was synthesized using a Sylgard 182 Silicone Elastomer Kit (Dow Chemical). Monomer (30 g) was mixed with the desired amount of ECM and stirred. Curing agent (3 g) was added and the mixture was stirred further before transferring to a Teflon mold and heated at T≈160°C for 8 min to activate the ECM expansion. The cast samples were then moved to room temperature and cured for 3 hrs.

*Tensile testing*
The stored energy was determined using a universal testing machine (Tinius Olsen 1ST) on samples prepared in the ASTM D638 Type IV dogbone shape, under standard strain rate. All samples were cast into ASTM D638 standard molds in the same fashion as described above in "*Synthesis of LCE*" and "*Making a foam with ECM*". They were loaded up to 500% strain, at a slow strain rate of 0.002 s$^{-1}$.

*Dynamic Mechanical Characterization*
Dynamic Mechanical Analysis was performed using a TA DMA 850 instrument under a constant frequency of 1 Hz. This method gives a good way to determine the glass and the nematic-isotropic transition points in the complex metamaterials materials where the usual Differential Scanning Calorimetry does not give a good signal. Scans were performed under temperature ramp mode at a rate of 2°C/min between -50°C and 150°C; all DMA runs were carried out under the same conditions to produce comparative results. Deconvolution analysis of DMA signals were conducted on the data interval between -20°C and 50°C.

*Impact (Drop test)*
The "drop tests" were performed using the PCB 208C05 force sensor, with the signal conditioning unit 482C05 added to process sub-millisecond signal of high amplitude. The force sensor was mounted on a heavy plate and a Perspex tube of inner diameter 55mm was mounted vertically above it to guide the dropping steel ball of mass *M*=564g (diameter 50mm). The length of the tube was exactly *H*=1m, so the mass was getting the speed $v = \sqrt{2gH} \approx$ 4.4 m/s, or the momentum $Mv \approx$ 2.5 N.s (the impact energy was 5.5 J, for reference). This momentum value was used to calibrate the force sensor, which exports the data in the units of Voltage vs. Time. The integral of the force peak over the time of collision has to be equal to the impulse of the collision. Since the platform is stationary, the impulse is $Mv$, giving the calibration constant α: that transforms the Voltage units into the Force units (in N):

$$Mv = \int F(t)dt = \alpha \cdot \int V(t)dt$$

We used the sharply defined impact peak of bare-metal collision, when the onset and the end of the impact are clear, to calculate this conversion constant, and then used it for all the other measurements.

*Polarized Microscopy*

Microscope photos characterizing the ECM distribution in LCE samples were taken using a Microscope, in bright field and between crossed polars. The human hair, measured to be 80 nm in diameter, was placed on sample surfaces as a scale reference, and to ensure the relative truth of scale in the images were taken so that both the hair and ECM were in focus.

**Acknowledgements**

This work was supported by the ERC H2020 Synergy grant 101167171 (ALCEMIST), and by Cambridge Smart Plastics Ltd.